# XSAT of linear CNF formulas

Bernd R. Schuh

Dr. Bernd Schuh, D-50968 Köln, Germany; bernd.schuh@netcologne.de

keywords: complexity, XSAT , exact linear formula, l-regularity, k-uniformity, NP-completeness

**Abstract.**

Open questions with respect to the computational complexity of linear CNF formulas in connection with regularity and uniformity are addressed. In particular it is proven that any *l*-regular monotone CNF formula is XSAT-unsatisfiable if its number of clauses *m* is not a multiple of *l*. For exact linear formulas one finds surprisingly that *l*-regularity implies *k*-uniformity, with *m* = 1 + *k(l-1)*) and allowed *k*-values obey *k*(*k*-1) = 0 (mod *l*). Then the computational complexity of the class of monotone exact linear and *l*-regular CNF formulas with respect to XSAT can be determined: XSAT-satisfiability is either trivial, if *m* is not a multiple of *l*, or it can be decided in sub-exponential time, namely $O(n^{\sqrt{n}})$ . Sub-exponential time behaviour for the wider class of regular and uniform linear CNF formulas can be shown for certain subclasses only.

**Introduction.**

The computational complexity of linear formulas has been studied extensively by Porschen et.al., see e.g. [1,2]. NP-completeness of exact satisfiability (XSAT) and not-all-equal satisfiability (NAE-SAT) was proven for linear monotone formulae and extended to exact linear formula (without monotony). Both variants of SAT were identified as NP-complete also for *l*-regular and even *k*-uniform subclasses of CNF formulas by these authors. Open questions remained concerning the NP-completeness of monotone *l*-regular and *k*-uniform exact linear CNF instances. Porschen et. al. conjectured that XSAT is NP-complete for this CNF class, and also in monotone *k*-uniform and *l*-regular linear CNF classes. This communication collects some properties of linear and exact linear CNF instances which might be of relevance to the cited conjectures. The paper is an enlarged version of earlier papers [3, 3'], in particular theorems 1,2,7, and 8.

Throughout this paper I will adopt the notation used in [1,2] which is shortly repeated here. A Boolean formula in Conjunctive normal form (CNF) by definition is a conjunction of clauses, where each clause is a disjunction of literals. A literal is an occurrence of a Boolean variable or its negation. In a <u>linear</u> CNF formula any two clauses have at most one variable in common. The class of such formulas is denoted by $LCNF$. In an <u>exact linear</u> formula any two clauses have <u>exactly</u> one variable in common. The class is denoted by $XLCNF$. A <u>monotone</u> formula contains positive literals only. Monotony is denoted by a subscript +, e.g. $LXCNF_+$. *l*-regularity is the property that each variable occurs *l* times. It is denoted by a superscript *l*, e.g. $CNF^l$. Finally, uniformity means that every clause contains the same number of literals. If this number is *k*, one writes e.g. $k\text{-}CNF$.

XSAT is the problem of deciding whether for a given CNF formula *F* there is a truth assignment (also called model) that evaluates exactly one literal in each clause of *F* to true. If there is at least one such assignment *F* is said to be x-sat, otherwise x-unsat, for short. If the context is clear we denote the number of clauses of some formula *F*, i.e. $|F|$ by *m*, its number of variables, $|V(F)|$, by *n*.

The paper is organized as follows.

First we consider exact linearity in general. This property imposes severe limitations on the structure of such formulas. Quite generally it can be shown that the majority of clauses in an exact linear CNF formula *F* is longer than the largest occurrence $l(x)$ of any variable $x$, i.e. $|V(C)| \geq \max_{x \in V(F)} l(x)$ for $m - \max_{x \in V(F)} l(x)$ many clauses $C \in F$. This is a generalization of the statement that *k*-uniform and *l*-regular XLCNF do not exist unless $k \geq l$ (as stated in theorem 13 in [2]). Some general relations for XLCNF formulas are then derived. When regularity is added further severe restrictions in exact linear formulas arise. In particular it is shown that an exact linear *l*-regular CNF formula necessarily is *k*-uniform, with certain restrictions on the allowed *k*-values. As a by-product we get the number of clauses and the number of variables for fixed *l* and *k* as $m = 1 + k(l-1)$ and $n = k^2 - k(k-1)/l$.

Then we turn to the question of XSAT-satisfiability of monotone formulas in general and introduce a simple method using some straightforward considerations concerning the number of true literals. Then we consider the implications of *l*-regularity for the x-satisfiability of general monotone CNF formulas. In particular it is proven that any monotone *l*-regular CNF formula is x-unsat if *m* is not a multiple of *l*. For monotone *l*-regular exact linear CNF formulas we get the interesting result that *F* either is x-unsat (if $k \neq 1 \pmod{l}$, i.e. *k*-1 is not divisible by *l*), or x-satisfiability is decidable in time less than $O(2^{\alpha \sqrt{n} \ln n})$ up to polynomial corrections, with some $\alpha$ depending on *l*. For monotone linear (not exact linear) formulas which are *k*-uniform and *l*-regular a similar type of sub-exponential behavior can only be found in restricted, rather artificial subclasses, namely those $F \in k - LCNF_+^l$ for which each clause is not connected to $l(l-2)$ other clauses, and a subclass we termed *entangled linear chains*.

Implications for NP-completeness of XSAT of linear and l-regular formulas is shortly discussed in a concluding section.

**Theorems and proofs**.

Throughout this section formulas and clauses are considered to be non-empty. Also a positive and negative literal of the same variable in one clause is excluded.

**Theorem 1:** Let $F \in XLCNF$ and let $L := \max_{x \in V(F)} l(x)$ be the largest occurrence of a variable in *F*. Then there are at most *L* clauses *C* in *F* with $|C| < L$.

PROOF: Let $C_1 \in F$ be a clause with $l(x_1) = L$ for some $x_1 \in V(C_1)$. Denote by $C_2, C_3, ..., C_L$ the other *L*-1 clauses that have $x_1$ in common with $C_1$. We assume $l(x) > 1$ in particular for the other variables from $V(C_1) \setminus \{x_1\}$. Then there exist

$$l(x_2) - 1 + l(x_3) - 1 + ... + l(x_{|V(C_1)|}) - 1 = \sum_{i=1}^{|V(C_1)|} l(x_i) - L - (|V(C_1)| - 1) =: m_{add}$$ many additional clauses

which have exactly one $x_i \in V(C_1)$ in common with $C_1$, but no variable from $V(C_1)$ in common with each other due to exact linearity. None of these clauses $C_{L+1}, C_{L+2}, ..., C_{L+m_{add}}$ can consist of less than *L* literals because they must have one variable in common with each of the *L* clauses $C_1, C_2, C_3, ..., C_L$. Therefore, for all these clauses $|V(C_{add})| \geq L$. Now assume that there exists a further clause. It should have a variable in common with $C_1$ due to exact linearity, but cannot have such a connection because all occurrences of the variables of $C_1$ are completed already. Thus no further clause exists. □

**Theorem 2**: For $F \in XLCNF$ the following relations hold

(i) The number of clauses $m = |F|$ is given by $m = 1 - |V(C)| + \sum_{x \in V(C)} l(x)$, for <u>arbitrary</u> $C \in F$.

(ii) $\sum_{x \in V(C)} l(x) - |V(C)| = \sum_{x \in V(C')} l(x) - |V(C')|$ for any two clauses C and C' of *F*.

PROOF: For $F \in XLCNF$ any clause *C* is connected to <u>all</u> other clauses by exactly one variable. Since every variable in *C* occurs *l* times in *F* the total number of connections of *C* to other clauses is $\sum_{x \in V(C)} (l(x) - 1)$. Since this must be the total number of clauses – 1 (the clause C itself) one gets the stated result for *m*.

Alternative Proof: From the considerations of the proof of theorem 1 it follows that the total number of clauses is given by $m = L + m_{add} = 1 - |V(C_1)| + \sum_{x \in V(C_1)} l(x)$. But all considerations stay valid, if one starts with an arbitrary $C \in F$. This proves (i).

(ii) follows from (i), since (i) shows that $\sum_{x \in V(C)} l(x) - |V(C)| = m - 1$ is an invariant independent of *C*. □

**Corollary 1**: For *k*-uniform XLCNF one has: The sum of occurrences of all variables belonging to a given clause $C \in F$ is a constant independent of *C*. This invariant is given by $m - 1 + k$.

PROOF: Follows from theorem 2, (i) and (ii) since $|V(C)| = |V(C')| = k$ for *k*-uniform formulas by definition. □

Together with theorem 1 this implies that *k* must be larger or equal to the largest occurrence in the formula. Otherwise exact linearity would be violated. Note that this is a generalization of the observation made in [2], that no *k*-uniform, *l*-regular XLCNF with *k* < *l* exist.

Now we turn to exact linear formulas which in addition are *l*-regular, i.e. each variable occurs exactly *l* times. The central result is the observation that *l*-regularity <u>implies</u> *k*-uniformity in exact linear formulas.

**Theorem 3**: The class $XLCNF^l$ consists solely of *k*-uniform formulas with either $k = 1 \pmod{l}$ or $m \neq 0 \pmod{l}$. Furthermore the number of clauses and variables of *F* is given by, respectively: $m = 1 + k(l-1)$, $n = k^2 - k(k-1)/l$.

PROOF: The first part of theorem 3 is a special case of theorem 2, with all occurrences being equal. Thus $m = 1 - |V(C)| + \sum_{x \in V(C)} l = 1 + |V(C)|(l-1)$. But *m* must be independent of *C*, so we have $|V(C)| = k$ independent of *C*, or $F = \emptyset$. For the second part of the theorem we observe that the

total number of literals of *F* can be counted in two different ways leading to $l|V(F)| = k|F|$ or $ln = km$ since *F* is *l*-regular and *k*-uniform (which basically is a special case of (iii) ). Thus $n = km/l = k^2 - k(k-1)/l$ . Since *n* must be an integer, $k(k-1)$ must be a multiple of $l$ for $F \neq \emptyset$ to exist. If $k-1 \equiv 0 \pmod{l}$ also $m \equiv 0 \pmod{l}$. All other allowed values of *k* lead to $m \not\equiv 0 \pmod{l}$. □

**Corollary 2**: For *l* prime the class $k - XLCNF^l$ splits into two subclasses with $k \equiv 1 \pmod{l}$ and $k \equiv 0 \pmod{l}$ respectively.

PROOF: Follows directly from theorem 3 which states that $k(k-1)$ must be a multiple of *l*. □

Note that the formula for *m* for block designs ( *k=l*) is a special case of a combinatorial consideration in a different context by Ryser [4].

Next I consider XSAT satisfiability of some monotone XLCNF classes. The method used will be as follows: For any XSAT-model *y* of *F*, i.e. a satisfying assignment which evaluates *F* to *true* with exactly one true literal per clause, the number of true literals must be equal to *m*, the number of clauses. Thus all XSAT-models of *F* are among the solutions of the equation $\sigma_F(y) = m$, where $\sigma_G(y)$ denotes the number of true literals of any CNF formula *G* under an assignment *y*. The solutions of the equation $\sigma_F(y) = m$ will be called pseudomodels, because all models are among these solutions but not vice versa. By counting pseudomodels one has an upper bound on the number of models. Each pseudomodel can be tested on XSAT-satisfiability in polynomial time. Since all models, if one exists, must be among the pseudomodels, the decidability of XSAT is bounded by the number of pseudomodels up to polynomial time corrections.

Now, for monotone CNF the total number of true literals is easy to calculate since each variable contributes either 0 or *l(x)* to the total sum: $\sigma_F(y) = \sum_{x \in V(F)} y_x l(x)$ , where $y_x \in \{0,1\}$ specifies the chosen assignment. We can now prove

**Theorem 4**: Let $F \in CNF_+^l$ and $|F| \not\equiv 0 \pmod{l}$ . Then *F* is x-unsat.

PROOF:

Let $\sigma_F(y)$ denote the total number of true literals in *F* for an arbitrary assignment *y*. The collection of pseudomodels is defined by $M_{pseudo}(F) := \{y \in \{0,1\}^n : \sigma_F(y) = \sum_{x \in V(F)} y_x l(x) = |F|\}$ . Since F is *l*- regular all $l(x)$ are equal, and consequently $\sigma_F$ can take values $l\nu$ with $\nu \in \mathbb{N}_n$, depending on the assignment. (To be precise: $\nu = |\{x \in V(F) : y_x = 1\}|$ ). Thus pseudomodels are defined by $\sigma_F(y) = |F| = l\nu$. But the number of clauses is not a multiple of $l$ by assumption. So the equation has no solution, $M_{pseudo}(F) = \emptyset$. There is no pseudomodel and consequently no model. □

Note that linearity behold exact linearity was not needed for theorem 4. A special case of theorem 4 is

**Theorem 5**: Let $F \in k - XLCNF_+^l$ and $k \neq 1 (\bmod l)$. Then $F$ is x-unsat.

PROOF: For $|F| = 1 + k(l-1)$ as proven in theorem 3, we have $|F| \neq 0 (\bmod l)$ since $k$-$1$ is not a multiple of $l$ by assumption. Thus $F$ is x-unsat according to theorem 4. □

Thus in order to clarify the computational complexity of the class of monotone $l$-regular exact linear CNF formulas one only needs to consider the subclass with $k = 1 (\bmod l)$. Note that Lemma 11 in [2] stating that all members of $k - XLCNF_+^k$ are x-unsat, is a special case of theorems 4 and 5 since $m = 1 + k(k-1)$ is not an integer multiple of $k$.

**Theorem 6:** Any $F \in XLCNF_+^l$ is either x-unsat or is decidable in time $O\left(\exp\left(f(l)\ln(n)\sqrt{n}\right)\right)$ or better for fixed $l$ and up to polynomial corrections, where $n = |V(F)|$.

PROOF: First we note that there is a $k$ and $F$ is $k$-uniform with $|F| = m = 1 + k(l-1)$, as proven in theorem 3. Now, if $k$-$1$ is not a multiple of $l$, $m$ is not divisible by $l$ and $F$ is x-unsat according to theorem 5. So, if by assumption, $k = 1(\bmod l)$ the equation $m = l\nu$ does have an integer solution $\nu$ $=m/l$. We can now proceed along the lines of the proof of theorem 4. Since $\sigma_F = 0$ for the assignment in which all variables are set to 0 (*false*) one can construct all pseudomodels corresponding to $\nu$ by flipping any $\nu$ of the $n$ variables from 0 to 1 (*true*). Thus

$$|M_{Fpseudo}| = \binom{n}{\nu} = \binom{n}{m/l} = \binom{n}{n/k}$$, where the last equality holds because $F$ is $l$-regular and $k$-uniform.

Since for $l$ and $k$ given $n$ and $m$ are uniquely determined by equations given in theorem 3, we can go to ever larger formulas, $n \longrightarrow \infty$, $l$ fixed, by choosing appropriately large $k(n)$. One finds $n/k \xrightarrow[n \to \infty]{} \sqrt{(l-1)/l}\sqrt{n}$. Now $\binom{n}{n/k}$ can be evaluated for large *n* (*l* fixed), e.g. by means of Stirlings formula, to give the stated result, with f(*l*) given by

$f(l) = \sqrt{(l-1)/l}\left(1 - \sqrt{(l-1)/l}\ln(\sqrt{(l-1)/l})\right)$. Finally, the worst algorithm to determine x-satisfiability of *F* would be to list all pseudomodels and check each of them for x-satisfiability (in time polynomial). Since any model, if one exists, is bound to be among the pseudomodels, the process of determining x-satisfiability of *F* is limited by their number, i.e. $|M_{Fpseudo}| \times p.t.$ many steps. □

*k*-uniform and *l*-regular linear CNF (not <u>exact</u> linear) are much less restricted, and their behaviour with respect to XSAT can be expected to differ from their exact counterparts considerably. Though all considerations concerning the pseudomodels with the result $|M_{Fpseudo}| = \binom{n}{\nu} = \binom{n}{m/l} = \binom{n}{n/k}$ stay valid, this time *n/k* stays O(*n*) because *k* and *l* can be chosen independently and need not be changed to attain ever larger formulas. Nonetheless one finds subclasses which still exhibit sub-exponential behaviour in the sense of theorem 6. To illustrate this point we construct an arbitrarily large *l*-regular *k*-uniform linear formula from its exact linear version.

**Definition**: Let F and G be two identical *l*-regular *k*-uniform exact linear CNF formulas except for their variable sets : $V(F) \cap V(G) = \emptyset$ . Then we call $F \cup G$ on $V(F) \cup V(G)$ a *linear 2-chain* and *F* its generator.

Obviously $F \cup G$ is an *l*-regular *k*-uniform linear formula which is twice as large , both in number of clauses and number of variables, as its generating formula. But since *F* and *G* are completely independent x-satisfiability can be determined by doing this for *F* alone and transferring the result to the other variable set.

A less trivial linear formula which cannot be divided into two non-overlapping subsets (non-overlapping means: for any pair of clauses of the two subsets their respective variable sets have no element in common) is an ***entangled linear 2-chain*** , defined as follows:

Let $F \cup G$ be a linear 2-chain and $l_{js}$ be its literals numbered with respect to the variable set $\{a_s\}, s \in \{1,...,n, n+1,...2n\}$ and clauses $C_j, j \in \{1,...m, m+1,...,2m\}$ . Then choose a clause number $j \in \{1,...m\}$ and a variable index $s \in \{1,...n\}$ , move $l_{js}$ from clause *j* in *F* to the same clause but the corresponding variable place in *G*, i.e. $j, s+n$ . To keep $F \cup G$ *l*-regular move $l_{j+m,s+n}$ to the corresponding place in *F*, i.e. $j+m, s$ . The resulting formula is l-regular k-uniform and linear by construction, but can no longer be divided into two non-overlapping subsets. We call it an entangled linear 2-chain. By adding further identical formulas on different variable sets and entangle them with the precursor one can generate arbitrarily long entangled linear chains, limited only by the number of literals of the generator. This way we have identified a non-trivial subclass of $k-LCNF^l$ which is sub-exponential with respect to XSAT:

**Theorem 7**: Let $F \in k-LCNF_+^l$ be a monotone entangled linear $p$-chain, $p \in \{2,...,\ln\}$. Then XSAT can be decided in sub-exponential time $O(\exp(\alpha \ln(n)\sqrt{n})$ at most.

Proof: Consider the unentangled $p$-chain first. X-satisfiability of the generator of the $p$-chain can be determined in sub-exponential time according to theorem 6. Since all assignments for the generator have the same effect on the other chain links when transferred accordingly, i.e. $y_{s+\lambda n} = y_s$ for $s \in \{1,...,n\}$ and $\lambda \in \{1,...,p-1\}$, complexity of the complete chain is only enhanced by a factor of $O(p)$. Now consider a specific entanglement between two chain links. Let $l_{js}$ be the literal that has been transferred to position $j, s+n$, and $l_{j+m, s+n}$ the one that is transferred backwards to $j+m, s$. If we set $y_s = y_{s+n}$ clause $j$ has the same number of true literals as clause $j+m$, and the situation is unchanged with respect to the unentangled 2-chain. The same consideration can be made for all other pairs of links in the chain. □

In the next theorem we introduce another subclass of $k-LCNF_+^l$ which has sub-exponential behaviour with respect to XSAT. In preparation we prove the following

**Lemma**: For $F \in k-LCNF^l$ let $\Delta = |\{(C,C') \in F \times F : V(C) \cap V(C') = \emptyset\}|/2$ denote the number of pairs of clauses which have no variable in common (double counting is taken care of by the factor 1/2), i.e. are not connected at all. And let $m_{XL} = 1+k(l-1)$ denote the number of clauses of the corresponding <u>exact</u> linear formula, and $m$ the number of clauses of $F$. Then $\Delta = m(m-m_{XL})/2$.

PROOF: Since in a linear formula any two clauses either have exactly one variable in common or none we have $m(m-1) = |\{(C,C') \in F \times F : V(C) \cap V(C') \in \{0,1\}\}|$

$$= |\{(C,C') \in F \times F : V(C) \cap V(C') = \emptyset\}| + |\{(C,C') \in F \times F : |V(C) \cap V(C')| = 1\}|$$

$$= 2\Delta + l(l-1)n.$$

For a *k*-uniform and *l*-regular formula $nl = mk$ additionally. Substituting *mk* for *nl and s*olving for $\Delta$ gives the stated formula. □

**Theorem 8**: For members of the class $k - LCNF_+^l$ for which each clause is unconnected to exactly $l(l-2)$ other clauses XSAT is decidable in time $O\left(\exp\left(\alpha \ln(n)\sqrt{n}\right)\right)$ or better up to polynomial time corrections.

PROOF: If *F* fulfills the assumptions of the theorem one can infer $\Delta = ml(l-2)/2$, with $m = |F|$. From the lemma we then have $m - 1 = l(l-2) + k(l-1)$ where $km$ has been substituted for $nl$ because *F* is *l*-regular and *k*-uniform. The relation for *m*-1 can be rewritten as $m + l = 1 + (k+l)(l-1)$. This is the formula for an <u>exact</u> linear formula which is *l*-regular and $k + l$-uniform and has *m+l* clauses. Indeed we can add *l* clauses to *F* with the following properties: they have *k+l* literals each and have one variable in common. So all in all there are $1 + l(k+l-1)$ new variables. Since each clause of the original *F* is unconnected to $l(l-2)$ clauses by assumption we can add *l* literals corresponding to *l* of the new variables to each of the original clauses in such a way that now each clause has exactly one variable in common with each other clause, the *l* new clauses included. Thus the newly constructed *F'* is an element of $(k+l) - XLCNF^l$. For the rest of the proof we must assume that *F'* is also monotone.

Now we need to show that the newly constructed $F'$ is x-sat iff $F$ is. Let $x_r$ be the variable that the *l* added clauses have in common. Assume that an assignment *y'* exists which makes $F'$ x-sat. This assignment must have $y'_r = 1$ and $y'_s = 0$ for all other new variables, otherwise the added clauses would not be x-sat. Now *F* remains, but since *y'* makes *F'* x-sat, also *F* must be x-sat. If on the other hand *y* is an assignment that x-satisfies *F*, then $y'_t = y_t$ for the old variables and $y'_r = 1$ and $y'_s = 0$ for the newly added ones is an assignment that x-satisfies *F'*.

According to theorem 6 x-satisfiability for *F'* can be determined in sub-exponential time. So the same holds for *F*. □

**Concluding remarks.**

Theorem 3 states a somewhat surprising result: *l*-regularity <u>implies</u> *k*-uniformity for exact linear formulas, and one cannot choose *k* and *l* arbitrarily. A tacit assumption in the proof of theorem 3 is that a formula $F \in XLCNF^l$ with at least *l* clauses exists at all. But it is not difficult to construct such formulas, at least for low values of *k* and *l*, e.g. $l = 3$, $k = 7$.

Porschen et. al. conjectured in [2] that NP-completeness of XSAT holds for $k - XLCNF_+$ and raised the question whether this property could even be transferred to $k - XLCNF_+^l$. Theorems 5 and 6 of this study give hints on how this question is to be settled. First of all, a large fraction of this class, namely those formulas where *k*-1 is not a multiple of *l*, turns out to be unsatisfiable with respect to XSAT, according to theorem 5. The rest, as shown in theorem 6, does not display the exponential behaviour of running times when tested for XSAT, as is usually expected for NP-class problems, $2^{\alpha n}$, but a faster efficiency <u>at most</u> of order $n^{\alpha \sqrt{n}}$. The results do not exclude the possibility that XSAT satisfiability of this class even is decidable in polynomial time. It would be interesting to find an appropriate algorithm.

Porschen et. al. also conjectured that $k-LCNF_+^l$ is NP-complete. In theorems 7 and 8 we have named subclasses of $k-LCNF_+^l$ which show the same sub-exponential behaviour as exact linear ones, i.e. a XSAT-decidability in less than $2^{cn}$ steps. On the other hand, these subsets are rather artificial and probably do not cover a substantial fraction of $k-LCNF_+^l$. So our study does not exclude the correctness of the cited conjecture.

Finally I would like to mention, that the method using pseudomodels (see proof of theorems 4 and 6) was applied to the monotone case here only. For non-monotone formulas a more general theorem [5] can be used and the results are more complicated and depend on the detailed structure of the distribution of positive and negative literals. E.g. one negative literal per variable is enough to destroy XSAT-satisfiability of instances from $CNF^l$ completely, because $\sigma_{min}$ is raised to *n* > *m*, so the necessary condition $\sigma = m$ cannot be achieved by any *F*. Thinking of *l*-regular exact linear HORN formulas as a further example, one has to deal with inhomogeneous distributions of positive literals in otherwise negative $XLCNF^l$. This leads to more complicated situations which are difficult to judge with respect to their asymptotic running times. An exception is a class of exact HORN formulas (XL and *l*-regular and exactly one positive literal per clause) with the additional condition that there is at most one positive literal per variable. Then $\sigma_{min} = m$, and there is exactly one pseudomodel to test, implying polynomial time behaviour, a situation which has been described in [6].